\documentclass[10pt,article,oneside]{memoir}
\pdfoutput=1

\date{}
\counterwithout{section}{chapter}

\usepackage{amsthm, amssymb, amsmath, latexsym, amsfonts, mathcomp}
\usepackage{xfrac}
\usepackage{verbatim}
\usepackage{color}
\usepackage{mathrsfs}
\usepackage{listings}
\usepackage{array}
\usepackage{url}
\usepackage{footnote}
\usepackage[style=authortitle-ibid,url=true,doi=false,isbn=false]{biblatex}
\usepackage{microtype}

\bibliography{Everything}
\usepackage{graphicx}

\newtheorem{proposition}{Proposition}[chapter]
\newtheorem{theorem}{Theorem}[chapter]
\newtheorem{lemma}{Lemma}[chapter]
\newtheorem{corollary}{Corollary}[chapter]
\newtheorem{definition}{Definition}[chapter]
\newtheorem{postulate}{Postulate}[chapter]
\newtheorem{axiom}{Axiom}[chapter]

\theoremstyle{definition}
\newtheorem{remark}{Remark}[chapter]

\def\dd{\mathrm{d}}
\def\Den{\mathrm{Den}}

\def\R{\mathbb{R}}

\def\M{\mathcal{M}}

\def\F{\mathcal{F}}  %

\newcommand{\dpp}[2]{\ensuremath{\frac{\partial #1}{\partial #2}}}

\providecommand{\red}[1]{ {\color{red} #1}}

\DeclareCiteCommand{\parencite}[\mkbibparens]
  {\usebibmacro{prenote}}
  {\usebibmacro{citeindex}%
    \clearfield{url}%
    \usebibmacro{cite}}
  {\multicitedelim}
  {\usebibmacro{cite:postnote}}
\DeclareCiteCommand{\footcite}[\mkbibfootnote]
  {\usebibmacro{prenote}}
  {\usebibmacro{citeindex}%
    \clearfield{url}%
    \usebibmacro{cite}}
  {\multicitedelim}
  {\usebibmacro{cite:postnote}}

\DeclareCiteCommand{\parencitenn}[\mkbibparens]
  {\usebibmacro{prenote}}
  {\usebibmacro{citeindex}%
   \usebibmacro{citetitle}}
  {\multicitedelim}
  {\usebibmacro{cite:postnote}}

\DeclareCiteCommand{\citenn}
  {\usebibmacro{prenote}}
  {\usebibmacro{citeindex}%
   \usebibmacro{citetitle}}
  {\multicitedelim}
  {\usebibmacro{cite:postnote}}

\let\oldmarginpar\marginpar
\renewcommand\marginpar[1]{\-\oldmarginpar[\raggedleft\tiny #1]%
{\raggedright\footnotesize #1}}

\title{A Primer on Differential Forms}
\author{\small{Christian Lessig\thanks{lessig@caltech.edu; this report resulted from my Ph.D. research performed at the Dynamics Graphics Project, University of Toronto}} \\ \small{Computing + Mathematical Sciences} \\ \small{California Institute of Technology}}

\begin{document}

\maketitle

\begin{abstract}
  This primer is intended as an introduction to differential forms, a central object in modern mathematical physics, for scientists and engineers.\footnote{This note is work in progress. Any feedback is highly appreciated.}
\end{abstract}

\section{Introduction}

Differential forms are ubiquitous in modern mathematical physics and their relevance for computations has increasingly been realized.
In the following, we provide a primer on differential forms with an emphasis on their relevance in modern classical mechanics which tries to convey the intuition underlying the concept.
A more detailed but still accessible discussion can be found in the book by Frankel,~\footcite{Frankel2003} and a rigorous treatment is available in the mathematics literature.\footcite{Marsden2004,Cartan2006,Novikov2006,Agricola2010}.
An introduction to a discrete formulation of differential forms, which provides an alternative perspective on the subject, can be found in a note by Desbrun and coworkers.\footcite{Desbrun2006}

\section{Why Differential Forms?}

Differential forms are central to the modern formulation of classical mechanics where manifolds and Lie groups are employed to describe the configuration and time evolution of mechanical systems.\footcite{Abraham1987,Arnold1989,Ratiu1999,Holm2009b}
One of the principal applications of differential forms in modern mechanics is the mathematical description of observables: infinitesimal measurements, which, when integrated, yield a value that can be verified through real world experiments, at least in principle.
Differential forms are thereby a natural choice when one requires that measurements satisfy:
\begin{enumerate}
  \item covariance, that is invariance under coordinate transformations;
  \item covariance under differentiation, which is crucial since the time evolution of most systems is described by differential equations;
  \item measurements are obtained by integration from the infinitesimal quantities employed to describe time evolution.
\end{enumerate}
The first requirement implies that differential forms have to be tensors, objects whose physical manifestation does not change under coordinate transformations, and the second requirement implies that these have to be anti-symmetric, leading to differential forms, anti-symmetric tensors that are ``\ldots ready (or designed, if you prefer) to be integrated''.\footcite{Desbrun2006}
Differential forms can hence be seen as a modern formulation of classical infinitesimals and, as we will see in the following, a formulation that adds much insight and efficacy to the concept.

Much of the utility of differential forms for calculations and the description of dynamical systems stems from exterior calculus, the calculus of differential or exterior forms, that provides the operators for working with forms, such as the wedge product and the exterior derivative.
Exterior calculus can thereby be understood as a generalization of vector calculus in $\R^3$.
In contrast to it, however, exterior calculus is defined on arbitrary dimensional, possibly curved manifolds, and even in $\R^3$ it elucidates much of the structure that is obfuscated in classical vector calculus.

\begin{figure}
  \begin{center}
    \includegraphics[trim = 45mm 67mm 35mm 55mm, clip, scale=0.6]{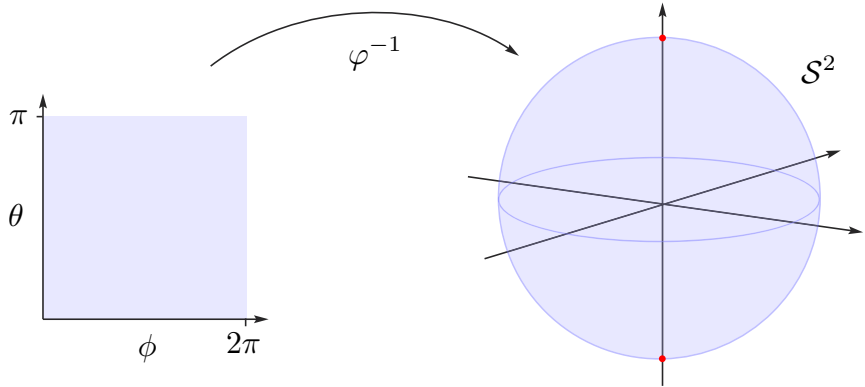}
  \end{center}
  \caption{The sphere $S^2$ as a two-dimensional manifold in $\R^3$. The chart $(U,\varphi)$ with $\varphi(U) = [0,2\pi] \times [0,\pi)$ covers $S^2$ up to a single point.}
  \label{fig:manifold:sphere}
  \vspace{-0.25in}
\end{figure}

The calculus of differential forms is not only vital to the mathematical description of mechanical systems in the continuum limit but, as began to be understood only recently, it is also crucial for numerical computations.\footcite{Desbrun2006,Arnold2010}
First applications were in electromagnetism.\footcite{Deschamps1981,Bossavit1997} but their relevance for many other systems has been demonstrated.\footcite{Elcott2007,Mullen2009a,Pavlov2011,Gawlik2011,DeWitt2012}.

\begin{remark}
  {\itshape
  In the following, we will need some ideas from manifold theory.
  For our purposes, it will suffice to think of a manifold $\M$ as an object that locally ``looks like'' Euclidean space $R^n$. 
  The local neighborhoods isomorphic to $\R^n$ are described by charts $(U,\varphi)$ which consists of an open set $U \subset \M$ and a chart map $\varphi : U \subset \M \to \varphi(U) \subset \R^n$, see Fig.~\ref{fig:manifold:sphere} for the interpretation of the sphere $S^2$ as a manifold in $\R^3$. 
  The inverse chart map $\varphi^{-1}$ is in this case given by 
  \begin{align*}
    \varphi^{-1} 
    =
    \left( \! \begin{array}{c}
      x \\ y \\ z 
    \end{array} \! \right)
    = 
    \left( \! \begin{array}{c}
      \sin{\theta} \, \cos{\phi}
      \\
      \sin{\theta} \, \sin{\phi}
      \\ 
      \cos{\theta}
    \end{array} \! \right)
    \, : \,
    \varphi(U) \to S^2
  \end{align*}
  so that, up to a point, $U = S^2$, and $\varphi{(S^2)} = [0,2\pi] \times [0,\pi)$.
  For simplicity, we will always assume in the following that the manifold of interest can be covered by a single chart.
}
\end{remark}

\section{Differential Forms in $\R^3$}

\begin{figure}
  \begin{center}
    \includegraphics[trim = 115mm 178mm 17mm 25mm, clip, scale=1.0]{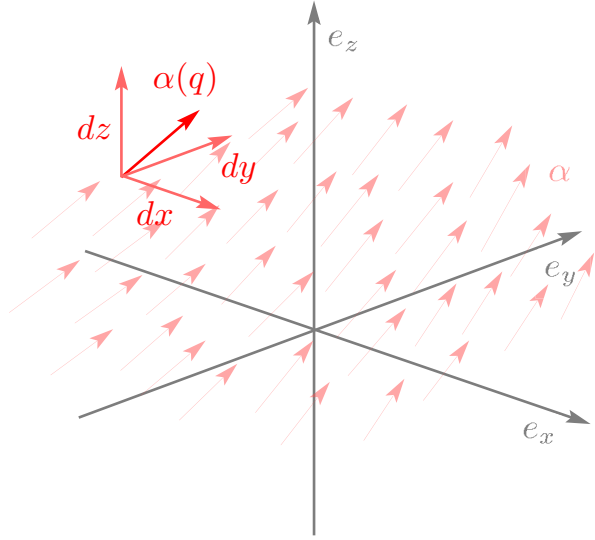}
  \end{center}
  \caption{A $1$-form in $\R^3$ can be thought of as a ``vector field'' with respect to the dual basis vectors $dx$, $dy$, $dz$.}
  \label{fig:one_form}
\end{figure}

Differential forms are naturally defined on manifolds, and this provides one of the most important advantages of the concept compared to more classical formulations such as vector calculus.
However, for pedagogic reason we will concentrate on differential forms in $\R^3$.
The essential features of forms are then still apparent but they are much more easily developed than in the general case.
We will begin by discussing $1$-forms, $2$-forms, and $3$-forms, and at the end of the section we will briefly comment on $0$-forms.

\paragraph{$1$-forms}
A $1$-form $\alpha \in \Omega^1(\R^3)$ can be thought of as a vector-valued object that is naturally integrated along a curve, a $1$-manifold in $\R^3$.
A physical example for the concept is a Newtonian force, which, when integrated along a curve, yields mechanical work.
As seen in Fig.~\ref{fig:one_form}, at a point $q \subset \R^3$ the $1$-form $\alpha(q)$ ``lives'' in the tangent space $T_q \R^3$ at the point, or more precisely in the cotangent space $T_q^* \R^3$, both of which can be identified with copies of $\R^3$ centered at the point $q$.
This is also apparent when we look at the coordinate expression for a $1$-form given by
\begin{align}
  \alpha(q) = \alpha_1(q) \, dx + \alpha_2(q) dy + \alpha_3(q) dz .
\end{align}
Similar to an ordinary vector, we can think of the $\alpha_i(q)$ as the projection of $\alpha(q)$ onto the dual basis vectors $dx,dy,dz$ that represent an infinitesimal integration in the $x$, $y$, and $z$ direction, respectively.
Hence, when we have a ``curve'' which is a straight line, say in the $x$ direction, then only the $\alpha_1$ component will contribute to the integral.
Formally, the dual basis vectors are defined through the biorthogonality condition
\begin{subequations}
\begin{align}
  & dx(e_x) = 1 \quad \quad dx(e_y) = 0 \quad \quad dx(e_z) = 0
  \\
  & dy(e_x) = 0 \quad \quad dy(e_y) = 1 \quad \quad dy(e_z) = 0
  \\
  & dz(e_x) = 0 \quad \quad dz(e_y) = 0 \quad \quad dz(e_z) = 1
\end{align}
\label{eq:diff_forms:basis_functions:biorthogonality}%
\end{subequations}%
where $e_x,e_y,e_z$ are the usual basis vectors for a vector in $\R^3$, or more precisely its tangent space $T_q \R^3$.

\begin{figure}
  \setlength{\abovecaptionskip}{-10pt}
  \begin{center}
    \includegraphics[trim = 20mm 185mm 20mm 20mm, clip, scale=0.65]{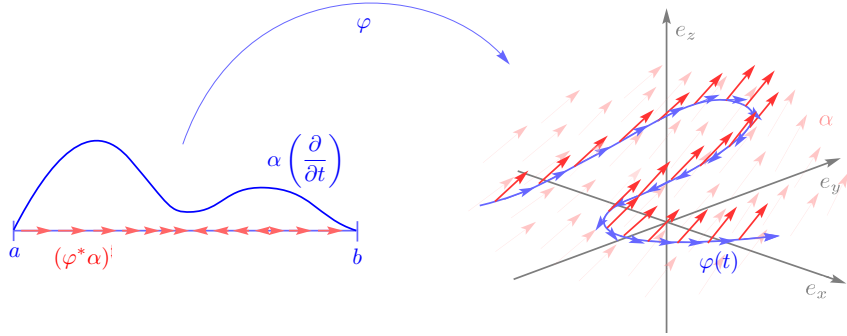}
  \end{center}
  \caption{A $1$-form $\alpha \in \Omega^1(\R^3)$ is an object that is naturally integrated along a curve. This is accomplished by pulling the form back from $\R^3$ onto the chart $[a,b]$ of the curve.}
  \label{fig:one_form_integration}
\end{figure}

Eq.~\ref{eq:diff_forms:basis_functions:biorthogonality} expresses a duality between vectors, which are linear combinations of the basis vectors $e_x,e_y,e_z$ and $1$-forms, which are linear combinations of the dual basis vectors $dx,dy,dz$.
This implies that a $1$-form, or covector, at a point is an object that naturally pairs with a vector to yield a real number. 
In mathematical jargon, a $1$-form is hence a functional over the space of vectors in $\R^3$.
For a vector field $\vec{A} : \R^3 \to \R^3$ whose element at $q \in \R^3$ is 
\begin{align}
  \vec{A} = A^1(q) e_x + A^2(q) e_y + A^3(q) e_z ,
\end{align}
the pairing with $\alpha \in \Omega^1(R^3)$ is determined by
\begin{align}
  \alpha(\vec{A})(q) 
  = \alpha \left( A^1 e_x + A^2 e_y + A^3 e_z \right)
\end{align}
where we omit for the moment the dependence of the components on $q$.
Also expanding $\alpha$, we have by linearity that
\begin{align*}
  \alpha(\vec{A})(q) 
  = \quad & \, \alpha_1 \, dx \left( A^1 e_x + A^2 e_y + A^3 e_z \right)
  \nonumber \\
+ & \, \alpha_2 dy \left( A^1 e_x + A^2 e_y + A^3 e_z \right)
\\
+ & \, \alpha_3 dz \left( A^1 e_x + A^2 e_y + A^3 e_z \right)
\nonumber
\end{align*}
and exploiting linearity once again yields
\begin{align*}
  \alpha(\vec{A})(q) 
  = \quad &\alpha_1 \, dx ( A^1 e_x ) + \alpha_1 \, dx ( A^2 e_y ) + \alpha_1 \, dx ( A^3 e_z )
  \nonumber \\
+ &\alpha_2 \, dy ( A^1 e_x ) + \alpha_2 \, dy ( A^2 e_y ) + \alpha_2 \, dy ( A^3 e_z )
\\
  + &\alpha_3 \, dz ( A^1 e_x ) + \alpha_3 \, dz ( A^2 e_y ) + \alpha_3 \, dz ( A^3 e_z )
\nonumber
\end{align*}
which is equivalent to
\begin{align*}
  \alpha(\vec{A})(q) 
  = \quad &\alpha_1 \, A^1 \, dx(e_x) + \alpha_1 \, A^2 \, dx(e_y) + \alpha_1 \, A^3 \, dx(e_z)
  \nonumber \\
+ &\alpha_2 \, A^1 \, dy(e_x) + \alpha_2 \, A^2 \, dy(e_y) + \alpha_2 \, A^3 \, dy(e_z)
\\
  + &\alpha_3 \, A^1 \, dz(e_x) + \alpha_3 \, A^2 \, dz(e_y) + \alpha_3 \, A^3 \, dz(e_z) .
\nonumber
\end{align*}
But by the biorthogonality condition in Eq.~\ref{eq:diff_forms:basis_functions:biorthogonality} we then have
\begin{align}
 \alpha(\vec{A})(q) &= \alpha_1 \, A^1 \, dx(e_x)
  + \alpha_2 \, A^2 \, dy(e_y)
  + \alpha_3 \, A^3 \, dz(e_z)
\end{align}
and hence
\begin{align}
  \alpha(\vec{A})(q) 
  = \alpha_1 \, A^1 + \alpha_2 \, A^2 + \alpha_3 \, A^3 .
 \label{eq:one_form:pairing}
\end{align}

\begin{remark}
  {\itshape
The dot product in the above equation is naturally expressed using the Einstein summation convention
\begin{align}
  \alpha(\vec{A})(q) 
  = \sum_{i=1}^3 \alpha_i \, A^i
  = \alpha_i \, A^i
\end{align}
which motivates also the choice of ``upstairs'' and ``downstairs'' indices for the components of vectors and covectors, respectively. 
  }
\end{remark}

\begin{remark}
  {\itshape
  Interestingly, in Eq.~\ref{eq:one_form:pairing} the dot product arises solely from the pairing of the $1$-form $\alpha \in \Omega^1(\R^3)$ with the vector $\vec{A}$ and is unrelated to the inner product 
\begin{align}
  \left\langle \vec{A} , \vec{B} \right\rangle = \vec{A} \cdot \vec{B} 
\end{align}
that provides a pairing of two vectors $\vec{A},\vec{B}$.
An inner product, when available, can be employed to simplify calculations with differential forms, but this is beyond the scope of this note.
}
\end{remark}

We mentioned before that $1$-forms are objects that are naturally integrated along a curve, while we just saw that they also naturally pair with a vector. 
How are these aspects related then?
A vector that is naturally defined for a curve $\varphi(t) : [a,b] \to \R^3$ is the tangent vector 
\begin{align*}
  \dpp{}{t}(t)
  \equiv \dpp{\varphi}{t}(t) 
  = X^1(t) e_x + X^2(t) e_y + X^3(t) e_z \in T_{\varphi(t)} \R^3
\end{align*}
which is an element in the tangent space $T_{\varphi(t)} \R^3$ of $\R^3$ at the point $q = \varphi(t)$ at which the curve is at time $t$.
To integrate the $1$-form $\alpha \in \Omega^1(\R^3)$ along the curve  $\varphi(t)$, we thus pair it at every point with the corresponding tangent vector $\partial / \partial t$.
The integral is hence given by
\begin{align}
  \int_a^b \alpha(\varphi(t)) \left( \dpp{}{t}(t) \right) dt
  = \int_a^b \alpha \left( \dpp{}{t} \right) dt .
\end{align}
The pairing in the above equation is equivalent to an operation known as pullback, 
\begin{align}
  \varphi^* \alpha : \Omega^1(\R^3) \to \Omega^1(\R^1),
\end{align}
that takes the $1$-form $\alpha \in \Omega^1(\R^3)$ on $\R^3$ to the $1$-form $\varphi^* \alpha \in \Omega^1(\R)$ on the real line, or more precisely to a $1$-form over the chart $[a,b]$ of the curve, see Fig.~\ref{fig:one_form_integration}.

\begin{remark}
  {\itshape 
  The pullback is useful not only for integration but it appears in many different contexts when one has a map between manifolds, or, as occurs often in applications, a map from a manifold onto itself.
  For example, in physical applications the finite time transport of an observable, represented by a differential form, is usually described by a pullback.
  }
\end{remark}

With the pullback, the integration of $\alpha$ can also be written as
\begin{align}
  \int_{\varphi(t)} \alpha 
  = \int_{a}^b \varphi^* \alpha
  = \int_a^b \alpha \left( \dpp{}{t} \right) dt 
  \label{eq:one_form:integration}
\end{align}
which shows that the pullback reduces the integration of a vector-valued $1$-form over an arbitrary curve in $\R^3$ to the integration of a scalar function over the real line, a well familiar operation for which for example the Riemann or Lebesgue integral can be employed.
With Eq.~\ref{eq:one_form:pairing} and the right-most formulation of the integral in Eq.~\ref{eq:one_form:integration}, we can obtain a coordinate expression for the integral given by
\begin{align}
  \int_a^b \alpha \left( \dpp{}{t} \right) dt
  = \int_a^b \vec{\alpha} \cdot \vec{X} \, dt 
  = \int_a^b \vec{\alpha} \cdot  dr ,
  \label{eq:line_integral:classical}
\end{align}
where $\vec{X} = (X^1,X^2,X^3)$ and $\vec{\alpha} = (\alpha_1,\alpha_2,\alpha_3)$.
Eq.~\ref{eq:line_integral:classical} is equivalent to the classical formula for the line integral of a vector field along a curve.
In fact, whenever one encounters a line integral, what one is integrating is a $1$-form, even if the computation is classically stated using vectors.
For example, any classical physics book will present a Newtonian force as a vector.\footcite[p. 1]{Goldstein2002}, but more correctly a force should always be considered as a $1$-form.\footcite[Chapter 7]{Arnold1989}.

\paragraph{$2$-form}
A $1$-form is naturally integrated over a $1$-manifold, a curve. 
It is then not hard to guess that a $2$-form is a vector-valued object that is naturally integrated over a $2$-manifold, a surface in $\R^3$. 
A physical example for a $2$-form is the fluid flux through a surface.
Analogous to $1$-forms, we can think of a $2$-form at a point also as an object that yields a real number when paired with two vectors. 

\begin{figure}
  \begin{center}
    \includegraphics[trim = 65mm 185mm 50mm 25mm, clip, scale=0.8]{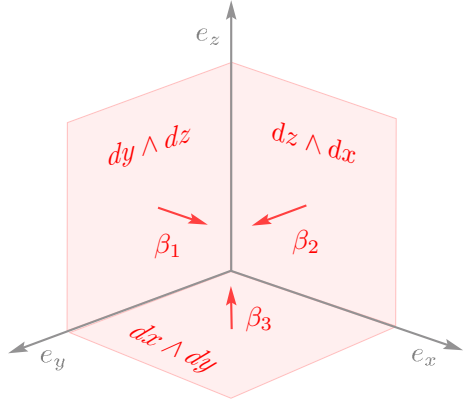}
  \end{center}
  \caption{The basis functions of a $2$-form $\beta \in \Omega^2(\R^3)$ can be interpreted as the flux through surfaces aligned with the coordinate system.}
  \label{fig:two_form_basis_functions}
\end{figure}

\begin{figure*}
  \setlength{\abovecaptionskip}{-10pt}
  \begin{center}
    \hspace{-0.1in}
    \includegraphics[trim = 25mm 170mm 15mm 25mm, clip, scale=0.7]{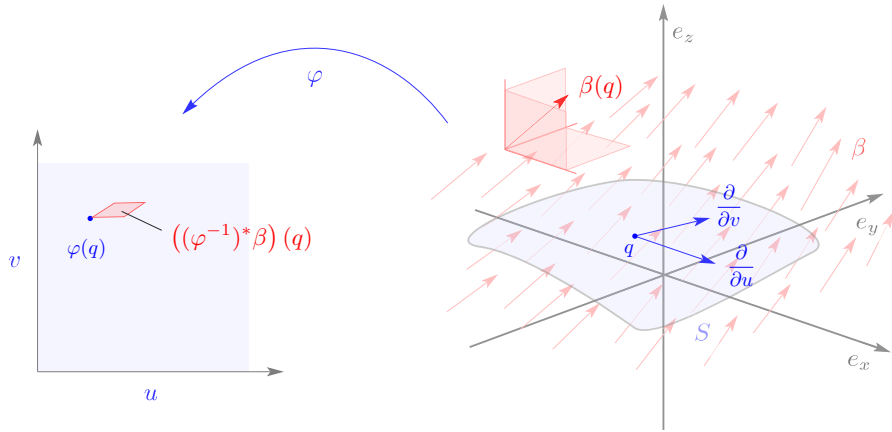}
  \end{center}
  \caption{For integration over a surface $S$, a $2$-form $\beta \in \Omega^2(\R^)$ is pulled back into a chart $(U,\varphi)$ of $S$ where it is a volume form over a Euclidean domain and can be integrated using standard integration techniques, such as Riemann or Lebesgue integration.}
  \label{fig:two_form_integration}
\end{figure*}

In components, a $2$-form $\beta \in \Omega^2(\R^3)$ is given by
\begin{align*}
  \beta(q) = \beta_1(q) \, dy \wedge dz + \beta_2(q) \, dz \wedge dx + \beta_3(q) \, dx \wedge dy .
\end{align*}
where the $dy \wedge dz$, $dz \wedge dx$, and $dx \wedge dy$ are elementary $2$-forms that span the space of all $2$-forms, they are the $2$-form basis functions.
These can be interpreted as infinitesimal fluxes through surfaces aligned with the coordinate axes, cf. Fig.~\ref{fig:two_form_basis_functions}, and a $2$-form hence represents the flux through a surface $S \subset \R^3$ with arbitrary orientation.
In analogy to $1$-forms, the flux through a concrete surface is thus obtained by the pullback $(\varphi^-1)^* \beta$, that is
\begin{align}
  \int_S \beta 
  = \int_{\varphi(S)} (\varphi^{-1} )^* \beta
  = \int_{\varphi(S)} \beta \left( \dpp{}{u} , \dpp{}{v} \right) du \, dv 
  \label{eq:two_form:integration}
\end{align}
where $(U,\varphi)$ is a chart for the manifold $S$, see again the example of a sphere $S^2$ in Fig.~\ref{fig:manifold:sphere}. 
As we see in Eq.~\ref{eq:two_form:integration}, the vectors naturally paired with a $2$-form for integration are the tangent vectors of a surface, see Fig.~\ref{fig:two_form_integration}.
Using a derivation analogous to those for differential $1$-forms, one can show that 
\begin{align}
  \int_{\varphi(S)} \beta \left( \dpp{}{u} , \dpp{}{v} \right) du \, dv 
  = \! \int_{\varphi(S)} \! \! B \cdot \vec{n} \, dA
  = \! \int_{\varphi(S)} \! B \cdot d\vec{A} ,
  \label{eq:two_form:integration:component}
\end{align}
where $B = (\beta_1,\beta_2,\beta_3)$, and the integration of a $2$-form hence corresponds to the classical integration of a vector field over a surface

\begin{remark}
  {\itshape
Differential $2$-forms in $\Omega^2(\R^3)$ are vector-valued, and hence when one only considers the components these appear identical to $1$-forms and ordinary vectors. 
However, the three are fundamentally different objects, and the difference is crucial from a physical point of view since they behave differently under a change of coordinates.
}
\end{remark}

\paragraph{3-forms}
With the previous discussions on $1$-forms and $2$-forms, the reader will be able to determine the principal properties of $3$-forms herself. 
Nonetheless, let us briefly summarize them.
A $3$-form $\gamma \in \Omega^3(\R^3)$ in $\R^3$ is an object that is naturally integrated over a volume $V \subset \R^3$, and pointwise paired with three vectors to yield a scalar.
In components, a $3$-form is given by
\begin{align}
  \gamma = \gamma(q) \, dx \wedge dy \wedge dz
\end{align}
where $dx \wedge dy \wedge dz$ is the basis function for the space of $3$-forms. 
Hence, a $3$-form is a scalar-valued object that is \emph{not} a scalar or function over $\R^3$, the crucial difference once again being the behaviour under a change of coordinates.
$3$-forms are the differential forms of maximal degree in $\R^3$---how should one integrate over a seven dimensional object in $\R^3$---and they are hence also known as volume forms.
One thus often writes $\Omega^3 = \Omega_{\textrm{vol}}$ to emphasize this.

In contrast to $1$-forms and $2$-forms, the integration of a volume form does not require a pullback.
The volume $V \subset \R^3$ is already a Euclidean domain, and hence integration can be performed directly so that
\begin{align}
  \int_V \gamma = \int_V \gamma(q) dq
\end{align}
where we also used the common short-hand notation $dq = dx \wedge dy \wedge dz$.

\begin{remark}
  {\itshape
The attentive reader will have noted that what is integrated is \emph{always} a volume form, and for $1$-forms in $\Omega^1(\R)$ and $2$-forms in $\Omega^2(\R^3)$ we had to employ the pullback to obtain a volume form on a lower dimensional manifold.
}
\end{remark}

\begin{remark}
  {\itshape
  Closely related to volume forms are densities, elements in $\Den(\M)$, that differ from elements in $\Omega_{\textrm{vol}}$ only by their behaviour when the orientation of the space changes, for example when one switches from a left-handed to a right-handed coordinate system. 
  For differential forms, when one integrates over a manifold with inverted orientation, then also the sign of the integral value changes. 
  For a $2$-form that is integrated over a surface in $\R^3$, for example, one obtains a negative flux when the direction of the normal is changed, cf. Eq.~\ref{eq:two_form:integration:component}. 
  However, when the flux represents energy or mass transport, then this change in sign is (in general) not meaningful.
  For densities the sign of an integral value does not change under a change of orientation, and these hence occur frequently in applications of differential forms in physics.
  }
\end{remark}

\paragraph{$0$-forms}

In the closing of this section, let us briefly mention $0$-forms. 
A zero dimensional manifold is a point, and hence ``integration'' of such forms amounts to evaluation: $0$-forms are functions, that is $\Omega^0(\R^3) \cong \F(\R^3)$.
This definition of zero forms is also natural in the light of the exterior complex that will be briefly discussed in the following.

\section{Exterior Calculus}

We have seen in the previous section that in $\R^3$ there are four different kinds of differential forms: $0$-forms, which correspond to functions, $1$-forms, which are integrated along curves, $2$-forms that are integrated over surfaces, and $3$-forms that are integrated over volumes.
This set of differential forms of different degree has a rich additional structure that in many applications provides crucial advantages over classical approaches.

A first connection between differential forms of different degree is provided by the wedge product that ``constructs'' a $(k+l)$-form from a $k$-form and an $l$-form, that is
\begin{align*}
  \wedge : \Omega^k \times \Omega^l \to \Omega^{k+l} 
\end{align*}
and which is the natural ``multiplication'' operator for differential forms, similar to the pointwise multiplication of polynomials where a polynomial of degree $k$ and a polynomial of degree $l$ also yield a polynomial of degree $k+l$.
A second connection is provided by the exterior derivative 
\begin{align*}
  \dd : \Omega^k \to \Omega^{k+1}
\end{align*}
which is a derivation that maps a $k$-form to a $(k+1)$-form. 
On $\R^3$ this yields the following sequence
\begin{align*}
  \Omega^0 
  \xrightarrow[]{\dd} \Omega^1
  \xrightarrow[]{\dd} \Omega^2
  \xrightarrow[]{\dd} \Omega^3
  \xrightarrow[]{\dd} 0
\end{align*}
where one also has $\dd \dd = 0$, that is the exterior derivative applied twice yields a trivial differential form.
Some intuition for the exterior derivative can be obtained by relating it to the classical operators of vector calculus when vectors and differential forms are identified using their components.
One then has
\begin{align*}
  \dd &: \Omega^0 \to \Omega^1 :: \nabla \quad \ \ \ \ \textrm{ (grad)}
  \\
  \dd &: \Omega^1 \to \Omega^2 :: \nabla \times \quad  \textrm{ (curl)}
  \\
  \dd &: \Omega^2 \to \Omega^3 :: \nabla \cdot \quad \ \ \textrm{ (div)}
\end{align*}
and $\dd \, \dd = 0$ corresponds to the classical laws that the curl of the gradient and the divergence of the curl vanishes.
Note that the use of differential forms clarifies when it is meaningful to apply curl or divergence to a ``vector field'', which is not apparent with classical vector calculus and can then only be deduced based on physical grounds, and with exterior calculus the three operations needed in the classical theory are all subsumed in one concept.

Much additional structure and powerful results are associated with differential forms:
\begin{itemize}
  \item de Rahm cohomology,
  \item interior product,
  \item Stokes therorem
  \item Cartan's formula
  \item \ldots
\end{itemize}
and these provide in practice a powerful language for working with differential forms.
However, of these aspects has to be deferred to another note and we refer to the literature cited at the beginning.

\subsection*{Acknowledgements}

Financial support by NSERC, GRAND, and NSF grant CCF-1011944 is gratefully acknowledged.

\newpage
\printbibliography[maxnames=20]

\end{document}